\documentclass[aps,prl,prabib,showpacs,nofootinbib]{revtex4}
\usepackage{graphicx} \usepackage{amsmath} \usepackage{amssymb}
\usepackage{amsfonts} \usepackage{bm}

\begin{document}

\newcommand{\be}{\begin{equation}} \newcommand{\ee}{\end{equation}}
\newcommand{\bea}{\begin{eqnarray}}\newcommand{\eea}{\end{eqnarray}}

\title{Normal mode analysis for scalar fields in BTZ black hole background} 

\author{Sayan K Chakrabarti} \email{sayan.chakrabarti@saha.ac.in}
\author{Pulak Ranjan Giri} \email{pulakranjan.giri@saha.ac.in} 
\author{Kumar S. Gupta} \email{kumars.gupta@saha.ac.in}

\affiliation{Theory Division, Saha Institute of Nuclear Physics, 1/AF
Bidhannagar, Calcutta 700064, India}

\begin{abstract}
We analyze the possibility of inequivalent boundary conditions for a
scalar field propagating in the BTZ black hole space-time. We find
that for certain ranges of the black hole parameters, the Klein-Gordon
operator admits a one-parameter family of self-adjoint extensions. For
this range, the BTZ space-time is not quantum mechanically
complete. We suggest a physically motivated method for determining the
spectra of the Klein-Gordon operator.
\end{abstract}


\date{\today}

\maketitle

\section{INTRODUCTION}

Probing a black hole space-time with scalar fields reveals important
information about thermodynamical properties of the system, including
Hawking radiation and black hole entropy
\cite{thooft1,sorkin,suss,satoh,thooft2,thooft3,ksg1,ksg2,ksg3}. In
this approach, the black hole geometry provides an external background
on which the scalar field dynamics is analyzed. There are however
several problems with such an analysis. The behaviour of the scalar
field at the event horizon may not be well behaved. The free energy of
the scalar field may show divergence due to infinite number of modes
contributing to it near the horizon \cite{thooft1}. Finally, for
certain geometries, the space-time may not be globally hyperbolic,
thereby leading to difficulties in the predicting the field
propagation \cite{isham,wald}. A possible way to address these issues is
by imposing suitable boundary conditions on the scalar fields. In
particular, the brick wall condition that the scalar field vanishes at
a certain distance away from the horizon leads to a well defined
prescription for the corresponding free energy and entropy
\cite{thooft1}. Such a boundary condition is by no means unique and
the physical results may depend on the particular choice made. It is
therefore a relevant exercise to classify all the possible boundary
conditions that lead to a well posed scalar field propagation and with
the view to ultimately find the effect of these different conditions
on various physical quantities of interest.

In this paper we shall study the possible choices of boundary
conditions and their consequences for the scalar field propagation in
the BTZ black hole background, which exists in 2+1 dimensions with a
negative cosmological constant \cite{btz}.  This system exhibits many
of the subtleties as mentioned above and yet, the scalar field
equation is exactly solvable in this geometry and the normal mode
analysis can be performed \cite{satoh, ken1, ken2}. This allows for
the possibility to analyze the effect of different boundary conditions
analytically, which is difficult for most other geometries. We shall
use the method of deficiency indices due to von Neumann \cite{reed} to
classify the various boundary conditions that can be imposed on the
scalar field. We find that for certain range of the system parameters,
there exists a one parameter family of self-adjoint extensions of the
corresponding Klein-Gordon equation. It often happens that the
determination of the domain using this technique leads to inequivalent
quantizations of the system \cite{jackiw,us1,us2,me1,me2}. Our analysis
will show that there are certain difficulties is adopting the usual
technique for the case of BTZ.

It has been argued that a space-time that is classically incomplete
may or may not be so in a quantum mechanical sense \cite{marolf}. The
particular prescription in \cite{marolf} proposes to define quantum
mechanical completeness in terms of the existence of a unique time
evolution of a scalar field probe. This ides is related to the issue of the quantum mechanical completeness of a differential operator \cite{reed}.
Our analysis clearly indicates that for certain range of
the system parameters, the time evolution of a scalar field in the BTZ
background is not unique and the corresponding space-time is not
quantum mechanically complete. In static geometries in which there is
a lack of global hyperbolicity, there exists a technique using
self-adjoint extensions \cite{wald} which leads to a well posed
problem. This technique, which has also been used to analyze quantum
singularities \cite{marolf}, is very similar to the one we use
here. However, we would like to emphasize that BTZ space-time with a
non-zero angular momentum is not static and hence the technique of
\cite{wald} cannot be used here to address the issue of global
non-hyperbolicity, which has to be addressed by the choice of suitable
boundary conditions at spatial infinity \cite{isham,satoh}.

This paper is organized as follows. In Section 2 we briefly review the
scalar field propagation problem in the BTZ space-time. In Section 3
we use von Neumann's method of self-adjoint extension to find the
deficiency indices of the radial part of the Klein-Gordon operator. In
Section 4 we discuss the implications of the self-adjoint extensions
and it physical relevance. Section 5 concludes the paper with some
discussions and an outlook.

\section{Scalar Field in BTZ black hole space-time}

We begin our discussion with the BTZ black hole, which is
$(2+1)$-dimensional space-time obtained from Einstein Equation with
negative cosmological constant $\Lambda = -\frac{1}{l^2}$. The
space-time is defined by the metric
\be \label{metric}
ds^2=- \left(-M + \frac{r^2}{l^2}+\frac{J}{4r^2}\right)dt^2+
\frac{dr^2}{\left(-M + \frac{r^2}{l^2}+\frac{J}{4r^2}\right)}
+r^2\left(-\frac{J}{2r^2}dt+d\phi\right)^2\,
\ee
where $M$ and $J$ denote the black hole mass and angular momentum
respectively.  For $0<|J|<M\ell$, there are two horizons, the outer
and inner horizons, corresponding respectively to $r=r_+$ and $r=r_-$,
where
\be \label{rpm}
r_\pm^2 =\frac {M\ell^2}2 \biggl\{ 1\pm \bigg[ 1 - \biggl(\frac
J{M\ell}\biggr)^2\biggr]^{\frac 12} \biggr\} \;.
\ee

The dynamics of a scalar field in this black hole space-time is given
by the Klein-Gordon equation
\begin{eqnarray}
\left(\square- \mu l^{-2}\right)\Psi(x) =0\,. \label{Klein}
\end{eqnarray}

This Klein-Gordon equation can be solved exactly. To see this, we
first use the separation of variables
\be
\Psi(x)= e^{-iEt}e^{im\theta}\mathcal{U}_E (r),
\ee
where $m \in Z$. Next we use the ansatz
\be
\mathcal{U}_E (r) = \left ( \frac{r^2}{l^2} - \frac{r_+^2}{l^2} \right )^{\alpha}
\left ( \frac{r^2}{l^2} - \frac{r_-^2}{l^2} \right )^{\beta} f(r) ,
\ee
where
\bea
\alpha &=& \frac{i l^2}{2(r_+^2 - r_-^2)} \left ( r_+ E - \frac{r_-}{l}n \right ) \nonumber \\
\beta &=& \frac{i l^2}{2(r_+^2 - r_-^2)} \left ( r_- E - \frac{r_+}{l}n \right ).
\eea
Finally, introducing the variable $z=\frac{r^2-r_-^2}{r_+^2-r_-^2}$,
the radial part of the scalar field equation becomes \cite{satoh}
\be \label{hyper}
H f(z) = 0,
\ee
where
\be \label{op} 
H = z(z-1)\frac{d^2 }{dz^2} + \{ c - (a+b+1)z \} \frac{d}{dz} - ab
\ee
and
\bea
a &=& \alpha + \beta + \frac{1}{2} (1 + \sqrt{1 + \mu}), \nonumber \\
b &=& \alpha + \beta + \frac{1}{2} (1 - \sqrt{1 + \mu}), \nonumber \\
c &=& 2 \beta + 1.
\eea
From (\ref{hyper}) and (\ref{op}) we see that the function
$f(z)$ satisfies the hypergeometric equation
\cite{as}. We shall restrict our attention to the analysis of the
Klein-Gordon equation in the region $[r_+, \infty)$.

The Klein-Gordon equation for the BTZ has been analyzed in detail
\cite{satoh, ken1, ken2}, which requires specification of the boundary
conditions. The BTZ black hole can be obtained by a discrete
quotienting of the universal covering space of $AdS^3$, which has a
timelike spatial infinity. This feature leads to the lack of global
hyperbolicity for the BTZ. Following \cite{isham}, this issue was handled in
\cite{satoh} by requiring that the solution of the Klein-Gordon
equation vanishes sufficiently rapidly at spatial infinity. We shall
also adopt the same boundary condition at spatial infinity. The
solution of (\ref{hyper}) that vanishes at the spatial infinity can be
analytically continued to the outer horizon $r_+$, where it diverges.
In order to regulate this divergence, it is customary to introduce a
brick wall parameter $\epsilon$ and require that the wave-function
vanishes at $x=z-1 = \epsilon$ , where $z=1$ is the outer horizon and 
$x$ is the near-horizon coordinate \cite{satoh}.
This set of boundary
conditions lead to a well posed problem with a finite solution as a
function of the brick wall cutoff. The corresponding thermodynamic
quantities can then be evaluated, which clearly depends on the
specific choice of the boundary conditions \cite{satoh}.

As stated earlier, our main purpose here is to investigate if other
boundary conditions exist which may also lead to sensible physics,
which is what we discuss in the next section.

\section{Inequivalent boundary conditions}

In this Section we shall try to find all possible boundary conditions
that lead to a sensible physical description for the scalar field
propagation in the BTZ background.  As mentioned earlier, the BTZ
space-time is not globally hyperbolic. In order to circumvent the
difficulties associated with the non global hyperbolicity, following
\cite{satoh}, we shall choose the solution of the wave equation
(\ref{hyper}) which vanishes at infinity. The boundary condition at
the brick wall is however not fixed by this criterion. Instead of
requiring the the solution vanishes at the brick wall, we shall demand
that the solution is square integrable near the outer horizon $r_+$
and that the corresponding Klein-Gordon operator is self-adjoint.  We
shall assume standard periodic boundary conditions on the angular
variables and investigate the possibility of any new boundary
conditions for the radial part of the Klein-Gordon operator, denoted
by ${\mathcal A}$.  We do this following von Neumann's method of
self-adjoint extensions \cite{reed}, whose basic features are recalled
below.

Let $T$ be an unbounded symmetric differential operator acting on a
Hilbert space ${\cal H}$ and let $D(T)$ be the domain of $T$. There
exists a criterion due to von Neumann for determining if $T$ is
self-adjoint in $D(T)$. For this purpose let us define the deficiency
subspaces $K_{\pm} \equiv {\rm Ker}(i \mp T^*)$ and the deficiency
indices $n_{\pm}(T) \equiv {\rm dim} [K_{\pm}]$. Then $T$ falls in one
of the following categories. i) $T$ is (essentially) self-adjoint iff
$( n_+ , n_- ) = (0,0)$. ii) $T$ has self-adjoint extensions iff $n_+
= n_-$. There is a one-to-one correspondence between self-adjoint
extensions of $T$ and unitary maps from $K_+$ into $K_-$. iii) If $n_+
\neq n_-$, then $T$ has no self-adjoint extensions.

In order to find the deficiency index $n_+$, we shall replace $E$ with
$i$ everywhere in ${\mathcal A}$ and find the square integrable
solutions of the corresponding equation. We are interested in the
physics in the region outside the outer horizon $r_+$. ${\mathcal A}$
is an unbounded differential operator defined in $ [r_+, \infty)$. It
is a symmetric operator in the domain $D({\mathcal A}) \equiv \{\phi
(x=\epsilon) = \phi^{\prime} (x=\epsilon) = 0,~
\phi,~ \phi^{\prime}~  {\rm absolutely~ continuous},~ \phi \in {\rm L}^2(\sqrt{g}dr)
\} $. We would next like to determine if ${\mathcal A}$ is self-adjoint in $D({\mathcal A})$, for which we proceed to find the deficiency indices of ${\mathcal A}$.

 Consider first the operator ${\mathcal A}$, in which the eigenvalues have been replaced with $+i$. Denote the resulting operator with ${\mathcal A}_+$. The equation which determines the deficiency index $n_+$ can then be written as
\be
{\mathcal A}_+ \phi_+ = 0,
\ee 
where $\phi_+$ is a function of $r$ (or equivalently of $z$ or $x$). We emphasize that the rhs of (10) is zero as the eigenvalues, which have been replaced with $+i$, already appear in $ {\mathcal A}_+$. The
solution of the corresponding radial equation which vanishes at
infinity is given by
\be \label{phi+}
\phi_+ = \left ( \frac{r^2}{l^2} - \frac{r_+^2}{l^2} \right )^{\alpha_+}
\left ( \frac{r^2}{l^2} - \frac{r_-^2}{l^2} \right )^{\beta_+} z^{-a_+} F(a_+, a_+ - c_+, a_+ - b_+ + 1, \frac{1}{z}),
\ee 
where
\bea
\alpha_+ &=& \frac{i l^2}{2(r_+^2 - r_-^2)} \left ( r_+ i - \frac{r_-}{l}n \right ), \nonumber \\
\beta_+ &=& \frac{i l^2}{2(r_+^2 - r_-^2)} \left ( r_- i - \frac{r_+}{l}n \right ).
\eea
and $a_+, b_+$ and $c_+$ are obtained by substituting $\alpha_+$ and
$\beta_+$ in eqn. (10).

The solution $\phi_+$ can be analytically continued near the outer
horizon where it takes the form
\be \label{hor}
\phi_+ \sim (z-1)^{\alpha_+} z^{\beta_+} e^{-i \xi_-} F(a_+, b_+, 2 \alpha_+ + 1, 1 - z) + A_+ e^{i \xi_+} (z-1)^{-\alpha_+} z^{-\beta_+} F (1-b_+, 1 - a_+, 1 - 2 \alpha_+, 1-z),
\ee
where
\be
A_+ e^{2 i \xi_+} = \frac{\Gamma(1-b_+)\Gamma(c_+ - b_+)\Gamma(a_+ +
b_+ - c_+)} {\Gamma(a_+)\Gamma(a_+ - c_+ + 1)\Gamma(c_+ - a_+ - b_+)}.
\ee
In terms of the near-horizon coordinate $x = z-1 $, the behaviour of
$\phi_+$ near the outer horizon defined by $z=1$ can be written as
\be \label{nh}
\phi_+ \sim e^{-i\xi_+} x^{\alpha_+} - A_+ e^{i \xi_+} x^{-\alpha_+}.
\ee 
In the near-horizon region, we therefore get that
\be
|\phi_+|^2 = x^{2 {\mathrm {Re}} (\alpha_+)} + A^2_+ x^{-2 {\mathrm
{Re}} (\alpha_+)} + A_+^2,
\ee
where ${\mathrm {Re}} (\alpha_+) = -\frac{r_+ l^2}{r^2_+ - r^2_-}$
denotes the real part of $\alpha_+$ in (12), which is a negative
number. Note that in the near horizon region, the measure $\sqrt{-g}
dr \sim r dr \sim dx$. Thus, the solution $\phi_+$ is square
integrable near the horizon if $0 > {\mathrm {Re}} (\alpha_+) >
-\frac{1}{2}$. Under this condition, we see that the deficiency index
$n_+ = 1$. A similar analysis shows that $n_- = 1$ as well. This in
the region $0 > {\mathrm {Re}} (\alpha_+) > -\frac{1}{2}$, the
operator ${\mathcal A}$ admits a one parameter family of self-adjoint
extensions, the latter being parametrized by a phase denoted by
$e^{i\gamma}, ~~ \gamma \in R$ (mod $2 \pi$).

 In the definition of the deficiency subspaces, instead of $\pm i$, in
 general one can consider an arbitrary complex number $\lambda$ and
 its complex conjugate, although it is usually sufficient to consider
 $\pm i$ alone. If we define the deficiency indices in terms of
 $\lambda$ and its complex conjugate, we find that the solutions
 spanning the deficiency subspaces are square-integrable near the
 horizon only with a band of values of the imaginary part of
 $\lambda$. This is an unusual situation for the BTZ space-time for
 which we do not have any definite interpretation.

\section{Implications of boundary conditions}

Application of the method of von Neumann has led us to the conclusion
that for certain range of the system parameters, the operator
${\mathcal A}$  is not self-adjoint but
admits a one parameter family of self-adjoint extensions. In addition,
using von Neuman's analysis \cite{reed}, it is possible to construct
the domain in which the operator ${\mathcal A}$ would be self-adjoint
when $0 > {\mathrm {Re}} (\alpha_+) > -\frac{1}{2}$. Since we have
$n_+ = n_- = 1$, the inequivalent domains are characterized by a $1
\otimes 1$ unitary matrix, which is just a phase denoted by $e^{i
\gamma}$. The domain $D_\gamma ({\mathcal A})$  in
which the operator ${\mathcal A}$ is self-adjoint is then given by
$D_\gamma ({\mathcal A}) = D({\mathcal A}) \oplus C \{ \phi_+ +
e^{i\gamma} \phi_- \}$, where $C$ is an arbitrary complex
constant. For each value of $\gamma \in [0, 2 \pi]$, we have a domain
$D_\gamma ({\mathcal A})$, which provides an inequivalent set of
boundary conditions compatible with the self-adjointness of the
Klein-Gordon operator.

In the usual problems of this type, the physical solution can
be made to belong to the domain of self-adjointness which leads to the
inequivalent quantizations of the system \cite{jackiw,
us1,us2,me1,me2}. For the case of the BTZ however this prescription for
the spectra does not seem to work. It can be shown that 
near the outer horizon a typical element of the domain $D_\gamma
({\mathcal A})$ has both oscillatory and non-oscillatory parts whereas
the physical solution is purely oscillatory \cite{satoh}. For this
reason we are not able to obtain the spectra corresponding to the
$D_\gamma ({\mathcal A})$ directly. We shall make some further
comments about this at the end of this section.

The issue of quantum mechanical completeness of the BTZ space-time can be addressed using the analysis presented here. We use the basic idea of
\cite{reed,marolf} where the quantum mechanical completeness of a given
geometry is related to the existence of a unique well defined quantum
time evolution of a scalar field probe in that background. This idea
can be formulated in terms of the self-adjoint extension of ${\mathcal
A}$. We have seen above that outside the parameter range $0 > {\mathrm
{Re}} (\alpha_+) > -\frac{1}{2}$, the deficiency indices $n_+ = n_- =
0$. This means that outside this range of parameters ${\mathcal A}$ is
essentially self-adjoint in $D({\mathcal A})$, which means that it has
a unique self-adjoint extension. Hence, outside the range $0 >
{\mathrm {Re}} (\alpha_+) > -\frac{1}{2}$, a scalar field in the BTZ
space-time has a unique time evolution and the corresponding geometry
is quantum mechanically complete. On the other hand, when $0 >
{\mathrm {Re}} (\alpha_+) > -\frac{1}{2}$, $n_+ = n_- = 1$ and the
Klein-Gordon operator admits a one parameter family of self-adjoint
extensions. The key issue is that within the context of the scalar
field dynamics in the BTZ background, there is no preferred choice
among the possible self-adjoint extensions. The corresponding dynamics
therefore suffers from this one parameter ambiguity, which cannot be
resolved within the context of this problem alone. This implies that
the BTZ space-time in not quantum mechanically complete when $0 >
{\mathrm {Re}} (\alpha_+) > -\frac{1}{2}$. In drawing this conclusion
we have assumed that the lack of global hyperbolicity for the BTZ can
be addressed by assuming a suitable boundary condition at infinity
\cite{isham, satoh}.

Finally, we shall attempt to give a proposal for finding alternate
boundary conditions at the outer horizon, from which a spectra
different from that of \cite{satoh} can be obtained. Our prescription
is not directly related to the domain of self-adjointness found above,
but is motivated by physical considerations \cite{wilczek}.

Let us first note that around the outer horizon $r_+$, given by $z=1$,
the two linearly independent expressions for $\mathcal{U}_E (z)$ are
given by $(z-1)^{\alpha}z^{\beta}F(1+b-c, 1+a-c, a+b+1-c, 1-z)$ and
$(z-1)^{-\alpha}z^{-\beta}F(1-a,1-b.c-a-b+1, 1-z)$ \cite{as}). As $z
\rightarrow 1$, the hypergeometric functions tend to $1$. Therefore a
general expression for $\mathcal{U}_E (z)$ near $z=1$ can be given by
\be
\mathcal{U}_E (z) \sim (z-1)^{\alpha}z^{\beta} + D (z-1)^{-\alpha}z^{-\beta}.
\ee
On the other hand, the solution $g$ which vanishes at infinity can be
analytically continued near the outer horizon and has the form
\cite{satoh}
\be
g (z) \sim (z-1)^{\alpha}z^{\beta} + e^{-2 i \pi \theta_0}
(z-1)^{-\alpha}z^{-\beta},
\ee
where
\be
e^{-2 i \pi \theta_0} = -\frac{\Gamma(1-b)\Gamma(c-b)\Gamma(a+b-c)}
{\Gamma(a)\Gamma(a-c+1)\Gamma(c-a-b)}.
\ee
These two expressions physically denote the same quantity, namely the
behaviour of the scalar field near the outer horizon, and they would
be compatible if $D$ is chosen as a pure phase $-e^{-2 i \pi
\delta}$. Furthermore, in order to regulate the wildly oscillatory
behaviour of the functions near the horizon, we introduce a brick wall
cutoff at $x = z-1 = \epsilon$. By comparing the two expressions near
the horizon we get
\be
\sin (\alpha \ln (\epsilon) + \pi \theta_0) =
\sin (\alpha \ln (\epsilon) + \pi \gamma).
\ee
A particular choice of the boundary condition could be that the
solutions vanishes at $x = \epsilon$, which will recover the results
of \cite{satoh}. However, instead of imposing a definite boundary
condition at a given point, we would also calculate the spectrum by
demanding that
\be
\delta = \theta_0 + 2 \pi k,
\ee
where $k$ is an integer. In the above equation, $\theta_0$ depends on
the system parameters as well as on the frequency, while $\delta \in
[0, 2\pi]$ is an arbitrary constant. For fixed values of the system
parameters, any choice of $\gamma$ would lead to a different spectrum,
which can be obtained numerically.

\section{Conclusions}

In this paper we have considered the possibility of inequivalent
boundary conditions for the radial part of the Klein-Gordon equation
of a scalar field in the BTZ background. Although the radial equation
in this case is exactly solvable, the analysis of self-adjoint
extension appears to be somewhat unusual. First, we have seen that the
solutions of the deficiency indices depend on the range of the
imaginary part of the complex quantity $\lambda$ that goes in the
calculation of the deficiency subspaces. This was certainly an
unexpected feature of this analysis. Second and even more surprising,
the domain that we obtain using von Neumann's analysis turns out to be
such that the physical solution does not appear to belong to it. Thus,
even though we found that the Klein-Gordon operator admits a one
parameter family of self-adjoint extensions, we could not obtain the
spectrum from the usual technique of von Neumann.

Due to this reason, we have proposed a way of obtaining the
inequivalent quantization of the spectrum motivated by a physical
approach towards the problem \cite{wilczek}.  In this approach, the
spectrum depends on an undetermined parameter and is independent of
any short distance cutoff. This is however somewhat misleading as the
equation determining the spectrum presupposes the existence of a brick
wall type cutoff. Thus, while in the usual analysis of \cite{satoh}
the dependence of the spectrum on the brick wall parameter is
explicit, in our general case, it is only implicit.

There is a proposal to classify the quantum completeness of space-time
using the self-adjoint extensions of the corresponding scalar field
equations \cite{marolf}. By that criterion, our analysis indicates
that the space-time associated with a rotating BTZ black hole is
quantum mechanically incomplete.

Finally, it appears likely that a renormalization technique used to
study certain singular potentials \cite{rajeev} can be used for this
geometry to obtain a beta function for the gravitational
coupling. Such an idea has already been used in the literature
\cite{suss}. In the usual spectrum of \cite{satoh}, if we demand that
the Newton's constant depends on the cutoff and demand that the
dependence is such that any particular energy level is independent of
the cutoff as the latter goes to zero, we would obtain a beta function
for the Newton's constant. In this case, it would be however difficult
to obtain an analytical expression due to the nature of the spectrum.

\end{document}